\documentclass[aps,twocolumn,groupedaddress]{revtex4}

\usepackage{graphics}
\usepackage{graphicx}

\def\fun#1#2{\lower3.6pt\vbox{\baselineskip0pt\lineskip.9pt
\ialign{$\mathsurround=0pt#1\hfil##\hfil$\crcr#2\crcr\sim\crcr}}}

\begin{document}

\title{Collinear Photon Exchange in the Beam Normal Polarization\\
Asymmetry of Elastic Electron-Proton Scattering}

\date{\today}
\author{Andrei V. Afanasev$^{a)}$ and N.P. Merenkov$^{b)}$}
\affiliation{
$^{(a)}$Jefferson Lab, Newport News, VA 23606, USA\\
$^{(b)}$ NSC ''Kharkov Institute of Physics and Technology'',\\
Kharkov 61108, Ukraine}

\begin{abstract}
The parity-conserving single-spin beam asymmetry of elastic electron-proton scattering
is induced by an absorptive part of the two-photon exchange amplitude. We demonstrate that this
asymmetry has logarithmic and double-logarithmic enhancement due to contributions of hard
collinear quasi-real photons. An optical theorem is used to evaluate the asymmetry
in terms of the total photoproduction cross section on the proton, predicting its
magnitude at a few  parts per million for high electon beam energies and small 
scattering angles. 
At fixed 4-momentum transfers, the asymmetry is rising logarithmically with increasing 
electron beam energy, following the high-energy diffractive behavior of total photoproduction 
cross section on the proton.

\end{abstract}

\maketitle

\section{Introduction}


It has been known for a long time \cite{ST1,Barut,Ru} that the two
photon exchange (TPE) mechanism can generate the single-spin
normal asymmetry (SSNA) of electron scattering due to a nonzero
imaginary part of the TPE amplitude $A_{2\gamma}$,
 \begin{equation}\label{1}
 A_n=\frac{2A_{Born}\Im (A^*_{2\gamma})}{|A_{Born}|^2} \,
 \end{equation}
where symbol $\Im$ denotes the imaginary (absorptive) part. The
one-photon-exchange amplitude $A_{Born}$ is purely real due to
time-reversal invariance of electromagnetic interactions.

The first calculations of the TPE effect on the proton \cite{aam}
predicted the magnitude of beam SSNA at the level of a few parts
per million (ppm). The effect appears to be small due to two
suppression factors combined: $\alpha=1/137$, since the effect is
higher-order in the electromagnetic interaction, and the electron
mass $m_e$ arising due to electron helicity flip. The predictions
of Ref.\cite{aam} which correspond only to elastic intermediate proton
state are in qualitative agreement with experimental data
from MIT/Bates \cite{Wells} and they were reproduced later in
Ref.\cite{Mark}.


However, the main theoretical problem in description of the TPE
amplitude on the proton is a large uncertainty in the contribution
of the inelastic hadronic intermediate states. In Ref.\cite{Mark}
the beam SSNA at large momentum transfers was estimated at the
level of one ppm, using the partonic picture developed in
Ref.\cite{Afanas} for TPE effects but not related to the electron
helicity flip.

Current experiments designed for parity-violating
electron scattering allow to measure the beam asymmetry with a
fraction of ppm accuracy \cite{Maas, E158,PViol} and may also
provide data on the parity-conserving beam SSNA. In fact, such
measurements are needed because beam SSNA is a source of
systematic corrections in the measurements of parity-violating
observables.

It was noted in \cite{aam} that while considering excitation of
inelastic intermediate hadronic states, the beam SSNA (Eq.(11) of
Ref.\cite{aam}), after factoring out the electron mass, has an
enhancement when at least one of the photons in the TPE loop
diagram is collinear to its parent electron. It is interesting
that this effect did not appear for the target SSNA. Similar
behaviour of the beam SSNA in the nucleon resonance region was
observed also in \cite{Pasquini} where authors used a
phenomenological model (MAID) for single-pion electroproduction.

In this paper we demonstrate that collinear photon exchange in the
TPE amplitude results in single- and double-logarithmic
enhancement of the beam SSNA, whereas such enhancement does not
take place for the target SSNA (with unpolarized electrons) and
spin correlations caused by longitudinal polarization of the
scattering electrons. For large electron energies and small
scattering angles, we use an optical theorem to relate the nucleon
Compton amplitude to the total photoproduction cross section and
obtain a simple analytic formula for the beam SSNA in this
kinematics.

\section{Leptonic and hadronic tensors}
\begin{figure}
\includegraphics[width=7cm]{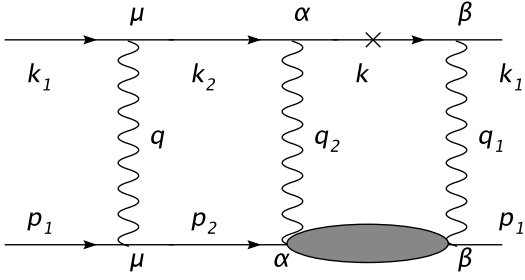}
\caption{Interference between the Born and the TPE box diagrams in elastic
e-p scattering that determines SSNA.}
\label{diagrams}
\end{figure}

First, we write the formula for SSNA in terms of rank-3 leptonic
and hadronic tensors which appear in the interference between the
Born and TPE amplitudes as shown in Fig.\ref{diagrams}.

\begin{equation}\label{2}
A_n=\frac{-i\alpha
Q^2}{\pi^2D(s,Q^2)}\int\frac{d^3k}{2E_k}\frac{L_{\mu\alpha\beta}
H_{\mu\alpha\beta}}{q_1^2 q_2^2} \ ,
\end{equation}
where $Q^2=-q^2$, $k (E_k)$ is the 3-momentum (energy) of the
intermediate on-mass-shell electron in the TPE box diagram, $q_1$ and
$q_2$ are the 4-momenta of the intermediate photons, $q_1-q_2=q$.
The factor $Q^2/D(s,Q^2)$ in Eq.(\ref{2})
is due to the squared Born amplitude, namely,

\begin{eqnarray}\label{3}
D(s,Q^2) &=& \frac{Q^4}{2}\big(F_1+F_2\big)^2 +[(s-M^2)^2-
\nonumber \\ &&
Q^2s]\Big(F_1^2+\frac{Q^2}{4M^2} F_2^2\Big ),
\end{eqnarray} 
 where $F_1(F_2)$
is the Dirac (Pauli) proton form factor, $M$ is the proton mass
and $s=(k_1+p_1)^2.$ Our sign convention for the beam asymmetry
follows from the definition of the normal vector with respect to
the electron scattering plane: ${\bf{k}_1}\times {\bf{k}_2}.$

Using the above notation, we have
\begin{equation}\label{4}
L_{\mu\alpha\beta}=\frac{1}{4}Tr(\hat{k}_2+m_e)\gamma_{\mu}(\hat{k}_1+m_e)(1-
\gamma_5\hat{\xi^e})\gamma_{\beta}(\hat{k}+m_e)\gamma_{\alpha}\ ,
\end{equation}
and
\begin{equation}\label{5}
H_{\mu\alpha\beta}=\frac{1}{4}Tr(\hat{p}_2+M)\Gamma_{\mu}(\hat{p}_1+M)(1-
\gamma_5\hat{\xi^p})\Im T_{\beta\alpha}\ ,
\end{equation}
where $m_e$ is the electron mass, $\xi^e(\xi^p)$ is the
polarization 4-vector of the electron beam (proton target),
$\Gamma_{\mu}=\gamma_{\mu}(F_1+F_2)-(p_{1\mu}+p_{2\mu})F_2/(2M)$,
and $T_{\beta\alpha}$ is in general  a non-forward proton Compton
tensor that describes any possible hadronic intermediate states in
the TPE amplitude. In accordance with Eq.(2) the single-spin
normal asymmetry probes the imaginary part of contraction of the
leptonic and hadronic tensors defined by Eqs.(\ref{4}) and
(\ref{5}), respectively. These tensors satisfy the conditions
\begin{eqnarray}\label{6}
&&q_{\mu}L_{\mu\alpha\beta} = q_{2\alpha}L_{\mu\alpha\beta} =
q_{1\beta}L_{\mu\alpha\beta} = 0,
\nonumber \\&&
q_{\mu}H_{\mu\alpha\beta} =q_{2\alpha} H_{\mu\alpha\beta} =
q_{1\beta}H_{\mu\alpha\beta} = 0,
\end{eqnarray}
separately for spin--independent and spin--dependent parts, as
follows from gauge invariance of electromagnetic interactions.

After some algebra we arrive at the following expression for the
model-independent leptonic tensor
\begin{equation}\label{7}
L_{\mu\alpha\beta}=L^{(un)}_{\mu\alpha\beta} +
L^{(pol)}_{\mu\alpha\beta} \ ,
\end{equation}
where the spin--independent part is
\begin{eqnarray}\label{8}
&&L^{(un)}_{\mu\alpha\beta}=
\frac{1}{2}q_1^2(g_{\mu\beta}k_{2\alpha}-g_{\mu\alpha}k_{2\beta})-
\frac{1}{2}q_2^2(g_{\mu\beta}k_{1\alpha}-g_{\mu\alpha}k_{1\beta})
\nonumber \\ &&
 -k_{\mu}[k_1k_2]_{\alpha\beta}+
\frac{1}{2}g_{\alpha\beta}(q_1^2k_{2\mu}+q_2^2k_{1\mu}-q^2k_{\mu})+
\\ &&
\frac{1}{2}q^2 (g_{\mu\alpha}k_{\beta}+g_{\mu\beta}k_{\alpha})
+k_{2\mu}(kk_1)_{\alpha\beta} +k_{1\mu}(kk_2)_{\alpha\beta}\ ,
\nonumber \\ &&
[ab]_{\alpha\beta} = a_{\alpha}b_{\beta} - a_{\beta}b_{\alpha}\
, \ \ (ab)_{\alpha\beta} = a_{\alpha}b_{\beta} +
a_{\beta}b_{\alpha}\ ,\nonumber
\end{eqnarray}
and the spin--dependent part is given by
\begin{eqnarray}\label{9}
&&L^{(pol)}_{\mu\alpha\beta}=im_e\bigg[-g_{\alpha\beta}(\mu q q_2
\xi^e) + k_{\beta}(\mu\alpha q \xi^e)+k_{\alpha}(\mu\beta q
\xi^e)
\nonumber \\ &&
+\xi^e_{\mu}(\alpha\beta q q_2) + (\xi^ek_2)(\mu\alpha\beta q_1) +
k_{2\mu}(\alpha\beta q_1 \xi^e)+
\nonumber \\ &&
k_{1\mu}(\alpha\beta q_2 \xi^e)+ \frac{1}{2}q^2(\mu\alpha\beta
\xi^e)\bigg]\ ,
\\ &&
(abcd)\equiv\epsilon_{\nu\lambda\rho\sigma}a_{\nu}b_{\lambda}c_{\rho}
d_{\sigma}\ , \nonumber
\end{eqnarray}
where the on-shell condition $k^2_{\mu}=m_e^2$ was used for
the intermediate electron 4-momentum.

If one of photon in the box diagram is collinear to its parent
electron, for example,
\begin{equation}\label{10}
q_1=xk_1, \ \ x= \frac{W^2-M^2}{s-M^2}\ ,
\end{equation}
where $W^2$ is the squared invariant mass of the intermediate
hadronic system, leptonic tensor can be written as
\begin{equation}\label{11}
L_{\mu\alpha\beta}=\frac{1-x}{x}q_{1\beta}L^B_{\mu\alpha} +
im_exL^{\xi}_{\mu\alpha\beta}\ .
\end{equation}
The tensor $L^B_{\mu\alpha}$ coincides with the Born one of elastic
electron--proton scattering and
\begin{eqnarray}\label{12}
&&L^{\xi}_{\mu\alpha\beta}=-g_{\alpha\beta}(\mu q k_1 \xi^e)
+\frac{q^2}{2}(\mu\alpha\beta \xi^e)+\xi^e_{\mu}(\alpha\beta q
k_1)+
\nonumber \\ &&
(\xi^ek_2)(\mu\alpha\beta k_1) + k_{2\mu}(\alpha\beta k_1 \xi^e)
+ k_{1\mu}(\alpha\beta k_2 \xi^e)\ . 
\end{eqnarray}
In the case of longitudinal polarization of the electron beam
$(\xi^e_{mu}=k_{1\mu}/m_e)$ the tensor $L^{\xi}_{\mu\alpha\beta}$ 
is zero. Therefore, we expect no contribution from considered
kinematics to the target SSNA or to longitudinal-spin correlations
because any gauge invariant hadronic tensor has to give zero after
contracting with $q_{1\beta}$ (see Eq.(6)).

In the case of the normal polarized electron beam
\begin{equation}\label{13}
\xi^e_{\mu}=\frac{2(\mu k_1 p_1 q)}{\sqrt{Q^2[(s-M^2)^2-Q^2s]}}\ .
\end{equation}
tensor $L^{\xi}_{\mu\alpha\beta}$ is not zero and the considered
collinear photon kinematics contributes with essential logarithmic
enhancement.

Therefore, conservation of the electromagnetic current that
follows from gauge invariance (Eq.(\ref{6})) is the reason why the
collinear intermediate photons appear in the TPE contribution to
the beam SSNA, but not to the target SSNA. By analogy, we do not
anticipate contributions from collinear-photon exchange in
unpolarized electron-proton scattering, parity-conserving and parity-violating
asymmetries due to longitudinal electron polarization
the normal polarization of leptons is not involved.

Let us consider the hadronic tensor. Small values of $Q^2$
correspond to the forward limit of nucleon virtual Compton
amplitude $T_{\beta\alpha}$. On the other hand, because $q_1^2$
and $q_2^2$ are also small in the collinear photon limit, we can
relate the forward Compton amplitude to the total photoproduction
cross section by real photons.

A general form of the Compton tensor $T_{\beta\alpha}$ in terms
of 18 independent invariant amplitudes that are free from
kinematical singularities and zeros was derived in
Ref.\cite{Tarrach}. Among these amplitudes we  choose the ones
that contribute at the limit $q^2\to 0$ and $q_1^2\to0.$ It
automatically constrains virtuality of the second photon to
$q_2^2\to0$.  There is only one structure that contains the tensor
$g_{\alpha\beta}$ and does not die off under the considered
conditions. It reads \cite{Tarrach}
\begin{eqnarray}\label{14}
&&T_{\beta\alpha}=\Big[-(\bar{p}\bar{q})^2g_{\alpha\beta}-
(q_1q_2)\bar{p}_{\alpha}\bar{p}_{\beta}
+(\bar{p}\bar{q})(\bar{p}_{\beta}q_{1\alpha} +
\nonumber \\ &&
\bar{p}_{\alpha}q_{2\beta})\Big] A(q_1^2,q_2^2,q^2,W^2) ,
\\ &&
\bar{p}=\frac{1}{2}(p_1+p_2), \ \ \bar{q} =
\frac{1}{2}(q_1+q_2)\ .\nonumber
\end{eqnarray}
It can be verified that
$T_{\beta\alpha}$ defined by the above equation satisfies the conditions
$T_{\beta\alpha}q_{1\beta}=T_{\beta\alpha}q_{2\alpha}=0.$

The normalization convention is chosen such that the imaginary
part of the quantity $(W^2-M^2-q_1q_2)^2A(W^2,q^2=0,q_1^2=q_2^2)$
is connected with the inelastic proton structure function
$W_1(W^2,q_1^2)$ by the following relation
\begin{equation}\label{15}
(W^2-M^2-q_1q_2)^2 \Im A(W^2,q^2=0,q_1^2=q_2^2)=
\end{equation}
$$\frac{\pi}{M}W_1(W^2,q_1^2)\ ,$$ and $W_1$, in turn, defines the
total photoproduction cross section \cite{Lev} as
\begin{equation}\label{16}
W_1(W^2,0)=\frac{W^2-M^2}{8\pi^2\alpha}\sigma^{\gamma
p}_{tot}(W^2)\ .
\end{equation}

Keeping in mind that the main contribution to the beam SSNA arises
from collinear photon kinematics, we can combine relations (5),
(14) and (15) and write hadronic tensor in the following form
$$H_{\mu\alpha\beta}=2\pi W_1\Big(F_1-\tau
F_2\Big)p_{1\mu}\Big(-g_{\alpha\beta}
-\frac{2 [p_1q]_{\alpha\beta}}{W^2-M^2}\Big)\ ,$$
\begin{equation}\label{17}
 \tau =\frac{Q^2}{4M^2}\ .
\end{equation}
When writing this last expression we neglect the terms
containing $(q_1q_2),$ which lead to the contribution of the order
$Q^2/W^2$ in the beam SSNA.

It may seem at first that in the considered limiting case of very
small $Q^2$ one can omit all terms proportional to $q$ in the hadronic
tensor. But such approximation is valid only for the symmetric
part of $T_{\beta\alpha}$ with respect to the indexes $\alpha$ and
$\beta.$ The reason is that the corresponding symmetric part of
the leptonic tensor (see Eq.(\ref{9})) contains the momentum
transfer $q$, and keeping it in the symmetric part of hadronic
tensor leads after contraction to additional small terms of the
order at least $Q^2/W^2.$ On the other hand, the antisymmetric
part of leptonic tensor contains terms which do not include the
momentum $q.$ Therefore, the antisymmetric part in Eq.(\ref{17})
has to be retained because it contributes at the same order with
respect to $Q^2/W^2.$ Note, however, that this antisymmetric part
of the hadronic tensor is not related to the polarized nucleon
structure functions, but it comes about as a consequence of the
gauge-invariant structure of Eq.(\ref{14}) even for a spinless
hadronic target.

\section{Master formula}

To compute the contraction of tensors in Eq.(2) we use the relations
$$-g_{\alpha\beta}p_{1\mu}L^{(pol)}_{\mu\alpha\beta}=2im_e[(p_1 q
q_1 \xi^e)+(k_1 p_1 q \xi^e)],$$
$$-[p_1q]_{\alpha\beta}p_{1\mu}L^{(pol)}_{\mu\alpha\beta} =
im_e(u-s)(p_1 q q_1 \xi^e),$$ $s+q^2+u=2M^2$ , which are valid for
the normal beam polarization
$((\xi^ek_1)=(\xi^ek_2)=(\xi^ep_1)=(\xi^ep_2)=0)$, and the
explicit form of 4--vector $\xi^e$ given by Eq.(\ref{13}). Then we
arrive at
\begin{eqnarray}\label{18}
&&L^{(pol)}_{\mu\alpha\beta}H_{\mu\alpha\beta}=\\ \nonumber
&&im_e\sqrt{Q^2}\Big(F_1-\tau
F_2\Big)\frac{(W^2-M^2)(s-W^2)} {4\pi\alpha}\sigma_T(W^2,q_1^2)\ ,
\end{eqnarray}
where $\sigma_T(W^2,q_1^2)$ is the total photoproduction cross
section with the transverse virtual photons. When integrating with
respect to $W^2$ we take
$\sigma_T(W^2,q_1^2)\rightarrow\sigma^{\gamma p}_{tot}(W^2)$ and
assume  $\sigma^{\gamma p}_{tot}(W^2)$ to be constant with energy
($\approx$ 0.1 mb, according to Ref.\cite{pdg}).

Taking into account Eqs. (2) and (18) one can write the beam SSNA
at small values of $Q^2$ as
\begin{equation}\label{19}
A_n^e=\frac{m_e\sqrt{Q^2}\sigma^{\gamma
p}_{tot}}{4\pi^3}\frac{F_1-\tau F_2}{F_1^2+\tau F_2^2}I\ ,
\end{equation}
$$I=\int\frac{d^3k}{2E_k}\frac{(W^2-M^2)(W^2-s)}{(s-M^2)^2}\frac{Q^2}{q_1^2
q_2^2}\ .$$ Here and further we use notation where $E_{k1},\ E_k
(\bf k_1, \ \bf k)$ are the energies (3-momenta) of the initial
and intermediate electron, respectively. The angular integration
in Eq. (19) can be done by introducing the Feynman parameter
$$\int\frac{d\Omega_k}{q_1^2q_2^2}= \int\limits_0^1
dy\int\frac{d\Omega_k}{[-2m_e^2+2(kk_y)]^2}\ , $$
$$k_y=yk_1+(1-y)k_2 = (E_{k1}; y{\bf{k}_1} + (1-y){\bf{k}_2}) , $$
$$ (kk_y)=E_{k1}E_k-2k|{\bf{k}_y}|\cos{\theta_y}, \ \
d\Omega_k=d\Phi d\cos{\theta_y}\ . $$

Integration over $d\Omega_k$ and Feynman parameter $y$ is
straightforward, leading to
\begin{equation}\label{20}
I = \frac{\pi}{2}\int_{m_{e}}^{E_{k1}}\frac{dE_k}{k}z(z-1) L, \
L= \frac{1}{K}\log\frac{2K+1}{2K-1}
\end{equation}
 where $z=E_k/E_{k1}$, and
$$\ L= \frac{1}{K}\log\frac{2K+1}{2K-1}\ ,
K=\sqrt{\frac{1}{4}+\eta}, \
\eta=\frac{m_e^2(E_{k1}-E_k)^2}{Q^2k^2}\ . $$
 We extended the upper
limit up to $E_{k1} $ because the difference between the value of
$E_k$ at inelastic threshold of pion production (when
$W^2=(m_{\pi}+M)^2$) and $E_{k1}$ is negligible at large $s.$

To calculate the integral in Eq.(\ref{20}), we note first that the
region where $k\simeq 0$ does not contribute because of the factor
of $L$. For this reason we can change integration with respect to
$E_k$ by integration over $k.$ Then we divide the integration
region into the following two parts, $ 0 < k < \lambda m_e$ and
$\lambda m_e< k < E_{k1}$, and choose the auxiliary parameter
$\lambda$ in such a way that
\begin{equation}\label{21}
\lambda >> 1, \ \lambda m_e<<E_{k1} << \sqrt{Q^2}\lambda \ .
\end{equation}

In the first region we can neglect $E_k$ as compared with $E_{k1}$
and write the corresponding contribution in the form
\begin{eqnarray}\label{22}
&&I_1=\pi\int\limits_0^{\lambda m_e}\frac{dk}{k}L \\ \nonumber 
&&=\frac{2m_e}{\sqrt{Q^2}}\int\limits_0^{\lambda\sqrt{Q^2}/2E_{k1}} \frac{2\pi
dz}{\sqrt{1+z^2}}\log(z+\sqrt{1+z^2})\\ \nonumber 
&&={\cal O}(\frac{2m_e}{\sqrt{Q^2}}).
\end{eqnarray}
We see that the contribution from the first region is negligible,
therefore we can choose zero for the lower limit of integration in $I_2$.
In the second region the quantity $\eta$ that enters $L$ is small
and we have
\begin{equation}\label{23}
I_2=\pi\int\limits_0^{1} dz (z-1)\Big(\log\frac{Q^2}{m_e^2}
+2\log\frac{z}{1-z}\Big)\ .
\end{equation}
The
integration in Eq.(\ref{23}) gives
$$I_2=\frac{\pi}{2}\Big(-\log\frac{Q^2}{m_e^2}+2\Big) \ .$$

 In the sum
$I_1+I_2$ we
arrive at the following master formula that defines the beam SSNA
for small values of $Q^2$ and takes into account contributions
from intermediate collinear photons in the TPE box diagrams
\begin{equation}\label{24}
A_n^e=\frac{m_e\sqrt{Q^2}\sigma^{\gamma
p}_{tot}}{8\pi^2}\frac{F_1-\tau F_2}{F_1^2+\tau F_2^2}
\Big(-\log\frac{Q^2}{m_e^2}+2\Big)\ .
\end{equation}
One can see that at fixed values of $Q^2$ the beam SSNA does not
depend on the beam energy if the total photoproduction cross
section is energy-independent. This remarkable property of
small-angle beam SSNA follows from unitarity of the scattering
matrix and does not rely on a specific model of nucleon structure.

\section{Numerical results and conclusion}

\begin{figure}[h]
\includegraphics[width=7cm]{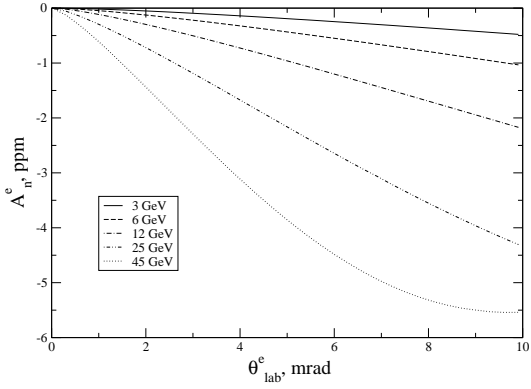}
\caption{Beam SSNA as a function of the lab scattering angle for
different beam energies: 3 GeV (solid line),
6 GeV (dashed line), 12 GeV (dash-dotted line), 25 GeV
(dash-double-dotted line) and 45 GeV (dotted line).}
\label{fig:thdep}
\end{figure}

The master formula for beam SSNA Eq.(\ref{24}) neglects possible
$Q^2$ dependence of the invariant form factor of the nucleon
Compton amplitude, which was taken in its forward limit during the
derivation. In numerical calculations, we estimate additional
$Q^2$ dependence by introducing an empirical form factor that was
measured experimentally in the Compton scattering on the nucleon
in the diffractive regime (see \cite{Bauer} for review). In the
following, we use an exponential suppression factor for the
nucleon Compton amplitude $\exp{(-B Q^2/2)}$, choosing the
parameter $B$=8 GeV$^{-2}$ that gives a good description of the
nucleon Compton cross section from the optical point to
$-t\approx$ 0.8 GeV$^2$ (see Table V of Ref.\cite{Bauer}). The
predictions of Eq.(\ref{24}) combined with the above described
exponential suppression are presented in Fig.\ref{fig:thdep} for
the electron scattering kinematics relevant for the E158
experiment at SLAC \cite{E158}.  We choose fit 1 of
Ref.\cite{block} for the total photoproduction cross section in
Eq.(\ref{24}). Exact numerical loop integration of Eq.(\ref{2})
and the analytic results of Eq.(\ref{24}) agree with each other
with accuracy better than 1\%. Contributions from the resonance
region ($W^2<$ 4 GeV$^2$) were estimated at 10-20\% at beam
energies of 3 GeV, but rapidly decreasing below 1\% at higher
energies. We also tested sensitivity of our results to $q_{1,2}^2$
dependence of the electroproduction structure function $W_1$
(Eq.(\ref{15})), taking various empirical parameterizations for
it.
 We found no sensitivity for SLAC E158 kinematics and only
moderate sensitivity ($\approx$ 10\%) when we extend our
calculation to lower energies ($\approx$3 GeV) and higher
$Q^2\approx$0.5 GeV$^2$. For beam energies of 45 GeV, numerical
integration shows that more than 95\% (80\%) of the result for
beam SSNA comes from the upper 1/2 (3/4) part of the
$W^2$-integration range. Based on the results of numerical
analysis, we conclude that the formula (\ref{24}) gives a good
description of beam SSNA at small $Q^2$ and large $s$ above the
resonance region.

\begin{figure}[h]
\vspace{0.1in}
\includegraphics[width=7cm]{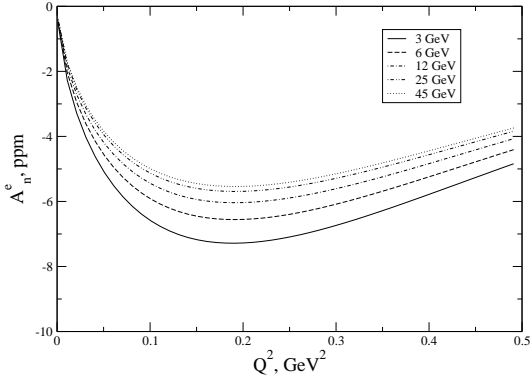}
\caption{Beam SSNA as a function of $Q^2$ for different beam energies. Notation is as in Fig.\ref{fig:thdep}.}
\label{fig:Q2dep}
\end{figure}
We also calculated the contribution of the elastic intermediate
proton state to the beam SSNA for high energies and small electron
scattering angles using the formalism of Ref.\cite{aam} and found
it to be highly supressed compared to the inelastic excitations.
For the kinematics of SLAC E158 \cite{E158}, this supression is a
few orders of magnitude due to different angular and energy
behavior of these contributions.

Shown in Fig.\ref{fig:Q2dep} are the calculations for beam SSNA as
a function of $Q^2$ for different energies of incident electrons.
One can see that at small $Q^2$, the asymmetry follows
$\sqrt{Q^2}$ behavior desribed by Eq.(\ref{24}), while at higher
$Q^2$ the asymmetry turns over and starts to decrease due to the
introduced exponential form factor $\exp{(-B Q^2/2)}$. It can be
seen that at fixed $Q^2$ the magnitude of beam SSNA is predicted
to be approximately constant, as follows from slow logarithmic
energy dependence of the total photoproduction cross section.

\begin{figure}[h]
\includegraphics[width=7cm]{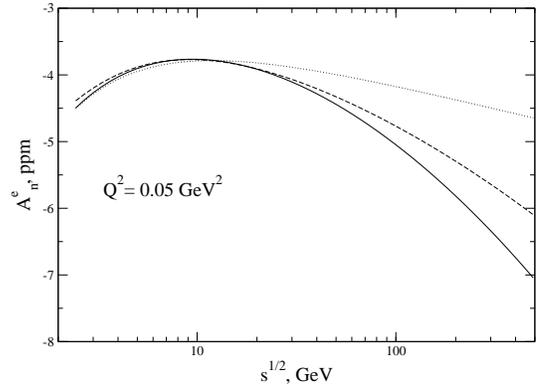}
\caption{Beam SSNA as a function of c.m.s. energy for fixed Q$^2$ = 0.05 GeV$^2$ for
different parameterizations of the total photoproduction cross section.
See Fig.\ref{fig:sigtot} for notation.}
\label{fig:sqsdep}
\end{figure}

\begin{figure}[h]
\vspace{0.2in}
\includegraphics[width=7cm]{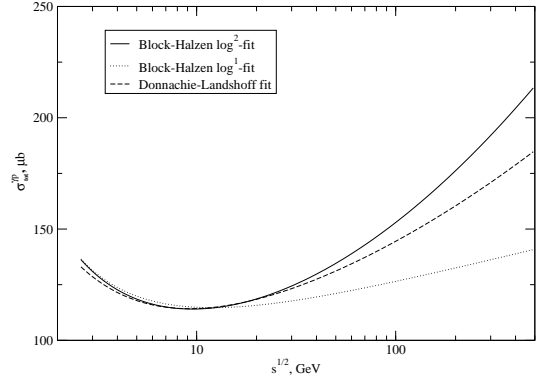}
\caption{Different parameterizations of total photoproduction cross section at high energies
used in the present calculation. A solid (dotted) line is a double-logarithmic fit 1
(single-logarithmic fit 3) from Block and Halzen
\cite{block}
and a dashed line is an original Donnachie and Landshoff fit \cite{DL}.}
\label{fig:sigtot}
\end{figure}

The latter feature is demonstrated in Fig.\ref{fig:sqsdep}, showing the calculated beam SSNA at fixed $Q^2$
in a wide energy range up to $\sqrt{s}$=500 GeV, where we used several parameterizations
for the total photoproduction cross section on a proton from Refs.\cite{block,DL}, shown in Fig.\ref{fig:sigtot}.
The physical reason for the almost constant photoproduction cross sections at high energies is believed to be soft
Pomeron exchange \cite{DL}, therefore the beam SSNA in the considered kinematics is sensitive to the physics
of soft diffraction.

The predicted $Q^2$ and energy dependence of beam SSNA, along with its relatively large
magnitude, is quite different from the model expectations assuming that no hadronic intermediate
states are excited in the TPE amplitude.  Our unitarity-based model of small-angle
electron scattering predicts the magnitude of the beam SSNA to reach a few ppm in a wide range of
beam energies. The good news is that it makes beam SSNA measurable with presently reached
fraction-of-ppm precision of parity-violating
electron scattering experiments \cite{PViol}. On the other hand, the experiments measuring
parity-violating observables need to use special care to avoid possible systematic uncertainties
due to the parity-conserving beam SSNA. Fortunately, these effects can be experimentally separated
using different azimuthal dependences of these asymmetries.

In the present paper we calculate the beam SSNA for small values
of $Q^2$ and provide physics arguments for the dominance of
contributions from collinear photons in the TPE mechanism. For
electron energies above the nucleon resonance region and small
$Q^2$ the contribution of collinear virtual photons leads to the
beam SSNA that is negative and has the order of
$m_e\sqrt{Q^2}\sigma^{\gamma p}_{tot}$, where $\sigma^{\gamma
p}_{tot}$ is the total photoproduction cross section on the
proton. This quantity is multiplied by the factor of the order
unity that includes a single-logarithm
term. The fact that the beam SSNA does not decrease with the beam
energy at fixed $Q^2$ makes it attractive for experimental studies
at higher energies, for example, the energies to be reached at
Jefferson Lab after the forthcoming 12-GeV upgrade of CEBAF.

\section*{Acknowledgements}

This work was supported by the US Department of Energy
under contract DE-AC05-84ER40150. N.M. acknowledges hospitality of Jefferson Lab,
where this work was completed.

\end{document}